\documentclass{aa}
\usepackage[varg]{txfonts}
\usepackage{natbib}
\usepackage{graphicx}
\usepackage{amsmath}
\usepackage{xcolor}
\usepackage[normalem]{ulem}
\usepackage{hyperref}
\bibpunct{(}{)}{;}{a}{}{,}

\begin{document}
\title{Swaying oscillations in Rayleigh-B\'enard convection\\ cast new light on  solar convection}

\author{F.\ Kupka\inst{1,2,3,4,5}
\and D.\ Fabbian \inst{2,3,5}
\and F.\ Zaussinger  \inst{6,7}
\and D.\ Kr\"uger \inst{1,2,5}
\and L.\ Gizon\inst{1,2,8} 
}
\institute{Max-Planck-Institut f\"ur Sonnensystemforschung, Justus-von-Liebig-Weg 3, 37077 G\"ottingen, Germany
\and Georg-August-Universit\"at G\"ottingen, Institut f\"ur Astrophysik und Geophysik, Friedrich-Hund-Platz 1, 37077 G\"ottingen, Germany
\and Faculty Comp. Sci. and Appl. Mathematics, Univ. of Applied Sciences, Technikum Wien, H\"ochst\"adtplatz 6, 1200 Wien, Austria
\and Wolfgang-Pauli-Institute c/o Faculty of Mathematics, University of Vienna, Oskar-Morgenstern-Platz 1, 1090 Wien, Austria
\and Fakult\"at f\"ur Mathematik, Universit\"at Wien, Oskar-Morgenstern-Platz 1, 1090 Wien, Austria
\and Brandenburg University of Technology Cottbus-Senftenberg, Department of Aerodynamics and Fluid Mechanics, Siemens-Halske-Ring 14, 03046 Cottbus, Germany
\and Hochschule Mittweida, Fakult\"at CB, Technikumplatz 17, 09648 Mittweida, Germany
\and 
Center for Astrophysics and Space Science, NYUAD Research Institute,  New York University Abu Dhabi, Abu Dhabi, UAE}

\date{Received xxx / Accepted xxx }
        
\abstract{Horizontally-periodic Boussinesq Rayleigh-B\'enard Convection (RBC)  is a simple model system to study the formation of 
large-scale structures in turbulent convective flows. We performed a suite of 2D numerical simulations of RBC between no-slip boundaries
at different Prandtl ($\mathrm{Pr}$) and Rayleigh ($\mathrm{Ra}$) numbers, such that their product is representative of the Sun's upper 
convection zone. When the fluid viscosity is sufficiently low  (${\mathrm{Pr} \lesssim 0.1}$) and turbulence is strong (${\mathrm{Ra} >10^6}$)
we find that large structures begin to couple in time and space. For ${\mathrm{Pr} = 0.01}$ we observe long-lived swaying oscillations of the upflows 
and downflows, which synchronize over multiple convection cells. This new regime of oscillatory convection may offer an interpretation
for the wave-like properties of the dominant scale of convection on the Sun (supergranulation).}

\keywords{convection --- hydrodynamics --- turbulence --- waves --- Sun: general}
\titlerunning{Swaying oscillations in Rayleigh-B\'enard convection}
\authorrunning{Kupka et al.}
\maketitle

\section{Introduction}

Solar supergranulation is the dominant scale of convection at the Sun's surface, with a characteristic size of $\sim30$~Mm 
and an $e$-folding lifetime of 1--3 days \citep{Rincon18b}. It has remained mysterious since its
discovery more than 60 years ago \citep{Leighton1962}.
In particular, the supergranulation pattern propagates with respect to the mean plasma flow and supports oscillations 
with a frequency of $1.8\ \mu$Hz or a period of about one week \citep{Gizon03a, Langfellner18a}. 
The oscillation frequency is a very robust feature of the observations and is measured to be independent of solar latitude, 
i.e.\ it is independent of the local rotation rate. No clear explanation has been proposed.

To explore the physical mechanisms behind the oscillation of the solar supergranulation pattern, 
we choose to consider the simplest simulation setup: Boussinesq RBC between two plates at rest with horizontally periodic boundary conditions.
Rather than starting from a fully realistic physical model, we deliberately focus on the simplest system that can potentially exhibit a “self-synchronization” of the largest energy-carrying convective scales, leading to large-scale oscillations similar to those observed.
We begin with a simplified 2D model without stratification or vertical shear, conditions that differ from those in the Sun. This configuration serves as a first step toward understanding the fundamental dynamics. In future work, we plan to progressively increase the model’s complexity by introducing vertical shear and stratification and extending the setup to three dimensions.
The 2D simulations presented in this paper, however, significantly extend the range of Rayleigh and Prandtl numbers used in traditional solar convection simulations, while keeping the computational cost under control.

Density stratification is a key ingredient of solar convection at supergranulation scales and is expected to influence the horizontal-to-vertical aspect ratio of 
supergranulation cells \citep[a point recognized early on by][]{Parker1973}. Observationally,  this aspect ratio is an open question and continues to challenge  
helioseismic analyses \citep[see, e.g.,][]{Hanson2024}.
While we acknowledge the importance of stratification, we choose to neglect it and we adopt the Boussinesq approximation. Instead, we focus on two other essential parameters of solar convection: the turbulence level and the viscosity of the solar plasma.

Semi-analytical studies almost always describe convection states near onset 
\citep{Bolton1986}. 
Numerical simulations are essential to study the hard turbulence regime. Particular care is needed 
in choosing the correct geometry, boundary conditions, and the physical parameters. 
The Rayleigh number $\mathrm{Ra}$ specifies  the strength of the thermal forcing and the Prandtl number 
$\mathrm{Pr}$ the strength of momentum diffusivity relative to heat diffusivity. 

Oscillations are known to exist in an RBC setup with walls, which are associated with a large-scale circulation (LSC) at 
very high Rayleigh numbers \citep{Krishnamurti81a,vanderPoel13a}. Several low-dimensional analytical models have been proposed 
to explain the coherent oscillations found in the LSC of turbulent RBC, 
including the ``mean wind reversal'' model \citep{Sreenivasan02a}.
A less constrained system consisting of a setup with walls where the bottom is cooled at the ends and heated in the center 
has recently been  studied with direct numerical simulations \citep{Reiter2020}. 
At low Prandtl numbers ($\mathrm{Pr} = 0.1$) and high Rayleigh numbers ($\mathrm{Ra} \gtrsim 3\times 10^8$), well defined oscillations 
are observed in the upflow lanes.
This result is encouraging. However, horizontal convection shows different heat and momentum transport scaling laws 
than convection driven by a temperature gradient aligned with the direction of gravity.

Simulations with horizontally periodic and stress-free boundary conditions at the top and bottom have been considered. 
No stable oscillatory state is found when the ratio of the horizontal to vertical domain size, $\Gamma$,  is too small \cite[e.g., $\Gamma=3$ in][]{Vincent99a}.
The importance of a large enough $\Gamma$ to allow the formation of large-scale structures in highly turbulent flows is evident in 3D simulations of Boussinesq RBC \citep{Parodi04a}.  
The condition $\Gamma \gtrsim5$ was shown to be sufficient for the horizontal extent of the structures to saturate at large $\mathrm{Ra}$ \citep{vonHardenberg08a,Stevens18a}.
For $\Gamma=5$, laboratory experiments with a liquid metal ($\mathrm{Pr}=0.03$) were carried out \citep{Akashi19a} and 
oscillations were observed when $\mathrm{Ra} > 10^5$; but these oscillations were attributed to a large scale circulation. 
We note however that for free-slip top and bottom boundary conditions and $\Gamma = 2$, a shearing mode instability 
may appear at sufficiently large $\mathrm{Ra}$ and moderate $\mathrm{Pr}$, such that  heat transport  occurs in bursts and the flow 
undergoes low-frequency global oscillations \citep{Goluskin2014}.

Solar convection calls for very low $\mathrm{Pr}$ and very high $\mathrm{Ra}$ \citep{Hanasoge2012}. So far, RBC simulations have been carried 
out at high $\mathrm{Ra}$ or low $\mathrm{Pr}$, but rarely in combination. Three-dimensional direct numerical simulations were carried out \citep{Pandey18a} 
with $\mathrm{Ra} = 10^5$ and  $0.005 < \mathrm{Pr} <70$, as well as  for $\mathrm{Ra}$ up to $10^7$ at a fixed  $\mathrm{Pr} = 0.7$, each with $\Gamma=25$ and side walls. 
The formation of superstructures were observed, but no oscillation was reported. 
Here we wish to use a geometrical setup that is more appropriate to study the solar problem and to cover a range of 
$\mathrm{Ra} \times \mathrm{Pr}$ (convective efficiency) that is closer to the solar value, while pushing down $\mathrm{Pr}$ and achieving long-duration time series. 
The aim is to reach a regime in which 2D RBC might support stable oscillations with periods of a few times 
the convective turnover time (i.e.\ solar frequencies in the microhertz range).

\section{Model and choice of physical parameters} 

The governing dynamical equations follow the Boussinesq approximation, which consists of assuming that the density stratification is small and 
the scalar variables fluctuate by a small amount about their mean values \citep{Lesieur08b}. We consider a 2D Cartesian box; the coordinates are denoted by $x$ and $z$ in the horizontal and upward directions respectively (the unit vector $\mathbf{\hat{z}}$ points upward). 
In our notation $\bf u$ is velocity, $T$ is  potential temperature,    
$\nu_{\rm kin}$ and $\kappa_T$ are the (constant)  momentum and heat diffusivities, $\alpha$ is the coefficient of volume expansion, and $g$ is the acceleration of gravity.
To write the dynamical equations using nondimensional variables, we scale all lengths by the height $H$ of the computational box, time by the diffusion time scale $\tau_{\rm diff} = H^2/\,\kappa_T$, and temperature by the temperature difference $\Delta T$  between the bottom and the top. 
Since the background state for Boussinesq RBC is hydrostatic, density does not explicitly appear in the equations. With the Prandtl number $\mathrm{Pr} = \nu_{\rm kin} / \kappa_T$ and the thermal Rayleigh number $\mathrm{Ra} =\alpha  g H^3 \Delta T/(\kappa_T \nu_{\rm kin})$, the governing equations are 
\begin{eqnarray}   
\nabla \cdot \bf u &=& 0, \\
{\partial \,{\bf u}}/{\partial t} + {\bf u} \cdot \nabla {\bf u}  
&=& - \nabla{p} + \mathrm{Pr} \nabla^2 {\bf u} +  \mathrm{Pr}\ \mathrm{Ra}\ T\ {\bf {\hat z}}, \\
{\partial \,T}/{\partial t} + {\bf u} \cdot \nabla T &=& \nabla^2 T,
\end{eqnarray} 
where $p$ is a pressure field.

We consider  horizontal periodic boundary conditions in the $x$-coordinate and impermeable boundary conditions at the top and bottom. 
For the top boundary ($z=H$) we have
\begin{equation}
u_z = u_x = 0  , \quad T=0,
\end{equation}
and for the bottom ($z=0$) we have
\begin{equation}
u_z =  u_x  = 0 
, \quad    T= 1.
\end{equation}
The no-slip boundary conditions avoid the development of a mean shear flow in the simulations.
The top and bottom boundary conditions lead to viscous boundary layers, which are smaller than the thermal boundary layers whenever $\mathrm{Pr} < 1$. 
We note that RBC in 2D domains show similar heat transport properties as in 3D \citep{vanderPoel13a,Schmalzl04a}, even at low Pr numbers \citep{Pandey2021-box}.

Our system is fully specified through $\mathrm{Ra}$, $\mathrm{Pr}$  and $\Gamma$. Stratification, sphericity and some other properties of the Sun's convection zone are not represented. In particular, the optically thin radiative transfer of the solar photosphere cannot be accounted for and sound waves are absent in Boussinesq RBC. The solar values of 
$\mathrm{Pr}$ and $\mathrm{Ra}$ are not achievable  with current computing resources \citep{Kupka17a}, but we carefully choose realistic values for 
$\mathrm{Ra}^* = \mathrm{Ra} \times \mathrm{Pr}$ (square of the ratio of the timescales of heat transport by diffusion and by buoyancy). 
In doing so, we relate the model to the upper layers of the solar convection zone where supergranulation takes place.
In this  series of experiments we take $\mathrm{Pr}$ in the range from  $0.01$ to $1$. From a solar structure model \citep{Spada2018} we can compute the buoyancy 
(Brunt-V\"ais\"al\"a) frequency averaged over the 
entire near-surface shear layer (upper 5\% by radius, i.e. $H=35$~Mm) to obtain the buoyancy timescale $\tau_{\rm buoy} = |N^2|^{-1/2} \approx 20$~hr. 
For the same solar model, we can read the diffusion timescale $\tau_{\rm diff} \approx 2.8\times 10^4\ {\rm hr} = 3.2$~yr averaged over the layer. With this, we find 
$\mathrm{Ra}^* = (\tau_{\rm diff}/\tau_{\rm buoy})^2 \approx 2.3\times 10^6$, 
and thus decide to carry out simulations covering the range $2 \times 10^{5} \le \mathrm{Ra}^* \le  2\times 10^{7}$ (the upper bound not being systematically 
investigated here yet due to its large computational cost).

We note that somewhat larger values might also be considered for Ra and thus
 larger values of $\mathrm{Ra}^*$, for instance, when taking the values from Table~2 of
 \citet{Ossendrijver2003} instead of values derived from the model of \citet{Spada2018}.
 However, any estimate of Ra is ultimately model-dependent, as it relies
 on the level of superadiabaticity, which remains largely unknown. 
 Therefore, we consider an order-of-magnitude estimate of the ratio 
 between diffusion and buoyancy time scales, derived from a standard 
 solar model, to be sufficiently accurate for the present work.

We set $\Gamma=8$, which accommodates 3 to 4 large-scale horizontal structures after relaxation. Based on our numerical experiments, this large value of $\Gamma$ is required to 
avoid the width of the simulation box to influence the dominant size of the horizontal structures. In dimensional units, we have $L = \Gamma  H = 280$~Mm, allowing for up to 
$\approx 8$ times the typical size of a solar supergranule. We also checked that the root mean square convective velocity $v_{\rm MLT} \sim  H  /(\sqrt{8}\ \tau_{\rm buoy}) 
\approx 170$~m/s from the mixing-length approximation is close to the observed root mean square horizontal velocity for solar supergranulation \citep{Rincon18b}. 
It is customary to define a convective turnover time $\tau_{\rm conv}=H/v_{\rm MLT}=57\ {\rm hr}$, which is approximately $0.002 \tau_{\rm diff}$. We note that both the estimated convective turnover and buoyancy times are relatively close to the observed lifetime of the solar  supergranulation pattern \citep{Rincon18b}. Thus, our experimental setup appears to be reasonable to study the questions raised in the introduction.

\section{Boussinesq convection experiments}

We use the \textsc{ANTARES} code \citep{Muthsam10a} to solve the problem numerically. This code has been used in the past to run convection experiments in the Boussinesq approximation \citep{Zaussinger_2011} and has been extensively tested \citep{Zaussinger_2013a, Kupka15a}. 
To study the regimes of interest of our model system and to understand the dependence of the results on the dimensionless numbers, we varied the Rayleigh and Prandtl numbers.  These values and the values of $\mathrm{Ra}^*$ are listed in Table~\ref{tab1}. 
The duration of the simulations are given in units of the thermal timescale. For all the cases we present here, this corresponds to  $\sim 250$ convective turnovers  with $\sim 11$ snapshots per buoyancy timescale.

We estimate the required spatial resolution of the Cartesian grid by means of scaling relations \citep{Grossmann00b}. 
We resolve the smallest structures (thermal boundary layer and Kolmogorov scale) with 3 to 4 grid points for $\mathrm{Pr} =1$ and 5 to 8 grid points for the other cases. We verify that the resolution is adequate by visualizing the solution in the boundary layers, and by comparing the physical viscosity given by $\mathrm{Pr}$ with the subgrid scale viscosity that would be used in a large-eddy simulation (LES) at the same resolution.

\begin{table*}[t]
 \centering
 \caption{Physical parameters of the various simulation runs.  \label{tab1}}
 \begin{tabular}{cccccc}
     number of grid points & $\mathrm{Ra}$ & $\mathrm{Ra}^* = \mathrm{Ra}\ \mathrm{Pr}$ & $\mathrm{Pr}$ & 
     time step  $[\tau_{\rm diff}]$
     & number of time steps used 
     \\
     &&&&& for Fourier analysis
     \\
    \hline\hline
    $960 \times 160$    & $2 \times 10^6$ & $2 \times 10^6$ & $1$        %
     & $4.0 \times 10^{-5}$ &  $7501$ \\
    $3000 \times 500$  & $2 \times 10^7$ & $2 \times 10^6$ & $0.1$     %
    &  $4.0 \times 10^{-5}$ & $7501$\\
    $3072 \times 512$  & $8 \times 10^6$ & $2 \times 10^5$ & $0.025$ %
    & $1.6 \times 10^{-4}$ & $1875$ \\ 
    $4500 \times 750$  & $2 \times 10^7$ & $2 \times 10^5$ & $0.01$   %
    &  $1.6 \times 10^{-4}$ & 1875 \\ 
 \hline
 \end{tabular}
\end{table*}

\section{Results}

\begin{figure}%
\centering
\includegraphics[width=0.95\columnwidth]{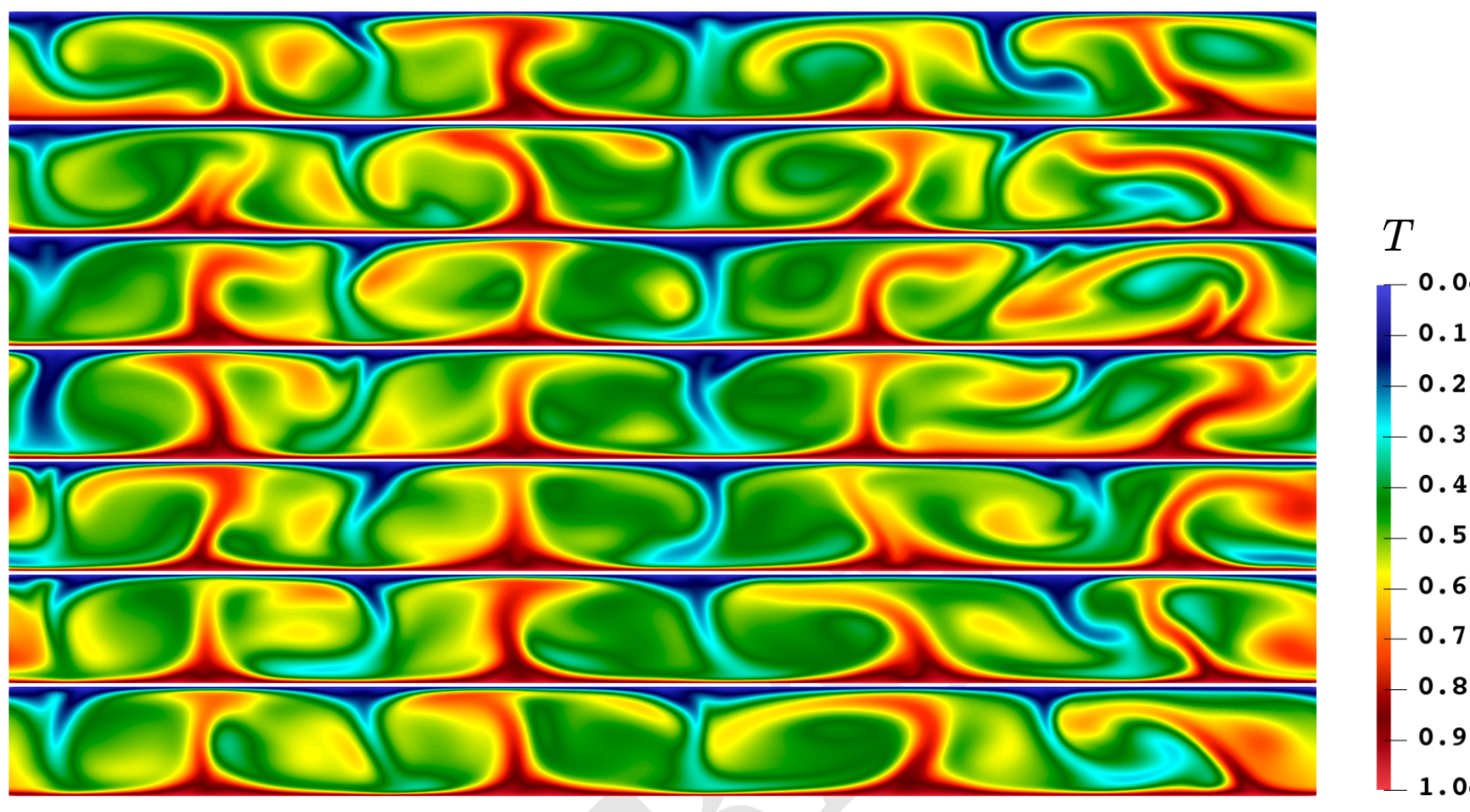}
\caption{Evolution of the temperature field for the simulation with $\mathrm{Pr}=0.01$ (last row in Table 1). 
The snapshots shown here are separated by 18 simulation time steps (time increases downward). 
The full sequence covers a little more than two convective turnover times and shows a full period of 
the ``swaying oscillations'' of the hot and cold  plumes. These oscillations are most easily seen in the movies 
on temperature (T-Pr001.mp4), horizontal velocity (u\_x-Pr001.mp4), and vertical velocity (u\_z-Pr001.mp4)
included as supplemental material and which are available \href{https://doi.org/10.17617/3.BHN9A4}{here}. In
all three movies the corresponding quantity is shown in dimensionless units: temperature is scaled
relative to the difference between bottom and top, velocities relative to the ratio between box size
and diffusion time scale.
\label{fig1}
} 
\end{figure}

The simulations provide us with long time series (over a year), which can be analysed to study the evolution of the system on  supergranulation time scales (multiple days). After an initial phase of plume formation and plume merging, a typical cell size is established. This first phase is completed after about 10 $\tau_{\rm conv}$. Then a further transition phase begins during which low-frequency stochastic power starts to build up. 
For $\mathrm{Pr} \gtrsim 0.1$, a third phase corresponding to a quasi-steady state is established after 20 $\tau_{\rm conv}$ or about  $0.04 \tau_{\rm diff}$. 

For $\mathrm{Pr} \lesssim 0.1$, we observe the development of an oscillatory behavior starting soon after the first transition phase (4 or 5 convective turnover times), which then persists throughout the entire simulation. The time evolution of the temperature field within the four main convection cells for $t>0.05\, \tau_{\rm diff}$ is shown in Fig.~\ref{fig1} for the case $\mathrm{Pr} =0.01$. The largest hot and cold plumes span the full vertical extent of the simulation domain. A large-scale synchronization of the horizontal motion of the hot plumes is clearly visible (see supplementary movie\footnote{Available at \href{https://doi.org/10.17617/3.BHN9A4}{https://doi.org/10.17617/3.BHN9A4}.}). 
The same phenomenon is observed for the cold plumes.  
The left and right swaying motion of the plumes is more prominent near the top boundary layer.

\begin{figure}[t]
\centering
\includegraphics[width=0.5\textwidth, trim=0.5cm 0 0.9cm 0, clip]{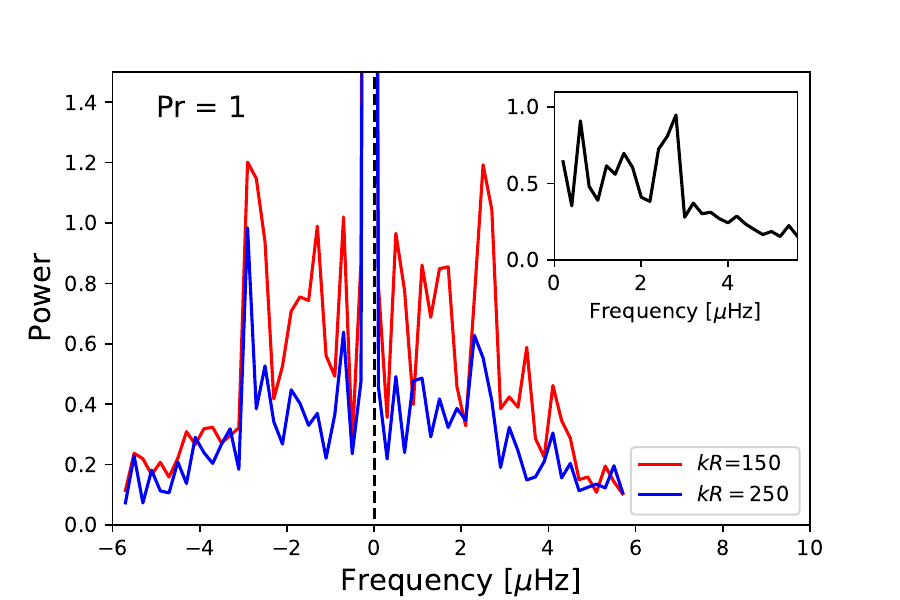}
\includegraphics[width=0.5\textwidth, trim=0.5cm 0 0.9cm 0, clip]{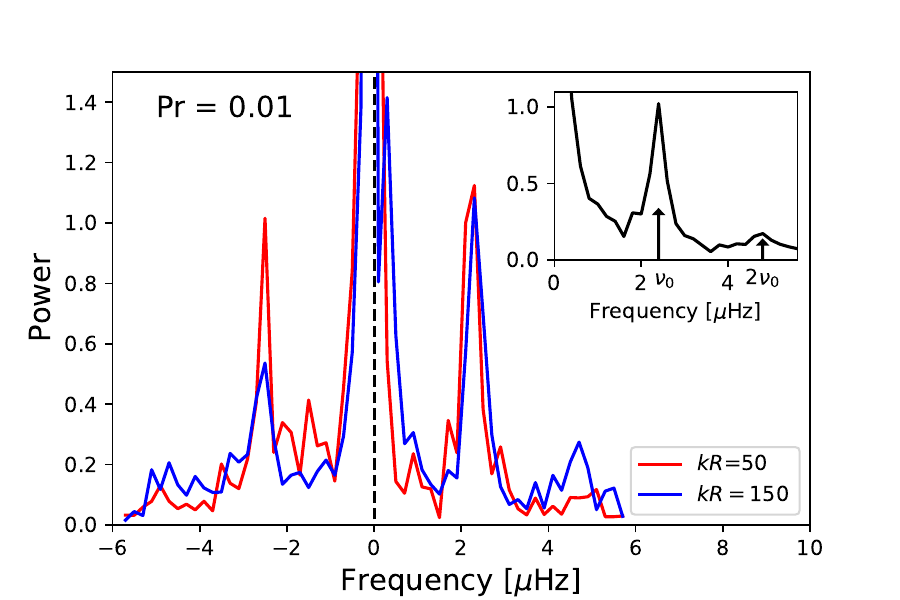}
\caption{      
     Power spectra of the horizontal velocity at selected values of the horizontal wavenumber  for $\mathrm{Pr}=1$ (top) and $\mathrm{Pr}=0.01$ (bottom) at fixed height $z_0=0.7H$. The insets show the symmetrized power averaged over $150 \le kR \le 250$ for $\mathrm{Pr}=1$, and averaged over $50 \le kR \le 150$ for $\mathrm{Pr}=0.01$. The arrows in the inset of the bottom panel point to the characteristic frequency $\nu_0=2.4\ \mu$Hz and to the first overtone at $2\nu_0$. Power is normalized with respect to maximum  power  (away from the central peak).
      \label{fig2}}
\end{figure}

In order to connect with the helioseismic observations of the supergranulation pattern \citep{Gizon03a}, we now turn to Fourier space. We compute the 2D power spectrum of the horizontal velocity $u_x$ at height $z_0=0.7H$ (below the top thermal boundary layer) for each numerical experiment:
\begin{equation}
    P(k,\nu) = \left| \iint u_x(x, z_0, t) e^{{\rm i}(kx-2\pi \nu t)}  {\rm d}x\ {\rm d}t  \right|^2 .
\end{equation} 
The data used in the analysis are restricted to the times after relaxation, i.e.\ $t > 0.05\ \tau_{\rm diff}$. All time series analysed in Fourier space have a duration of $0.3\ \tau_{\rm diff}\approx 1$~year (the exact number of corresponding time steps is given in the last column of Table~1). This implies a frequency resolution of $0.03\ \mu$Hz. Depending on the experiment, the Nyquist frequency is either $116\ \mu$Hz or $29\ \mu$Hz, i.e. much larger than required to cover the frequency range of the solar observations. In order to reduce the level of stochastic noise, the power spectrum may safely be binned down to lower frequency and spatial resolutions.

Figure~\ref{fig2} shows the power spectrum as a function of frequency $\nu$ at two particular wavenumbers, $k=150/R$ and $k=250/R$ where $R=696$~Mm is the solar radius (for comparison, $k=120/R$ is a typical value for supergranulation).
For the $\mathrm{Pr}=1$ simulation, we notice significant convective power below $6\ \mu$Hz, with enhanced power below $3\ \mu$Hz.
The peak at zero frequency indicates that, unlike $u_z$, the time average of $u_x$ does not vanish, but it remains very small (of order  $\sim 1$ m/s and depends on depth). 
As the Prandtl number drops to $\mathrm{Pr}=0.01$, we observe two very prominent peaks of excess power at frequencies $\pm 2.4\ \mu$Hz. The power at positive and negative frequencies is of the same order, with perhaps a noticeable asymmetry when $kR=150$ (the pattern is allowed to travel horizontally). 
It is remarkable that, without further tuning the parameters of the system, we find an oscillation frequency that is very close to that of the observations ($1.8\ \mu$Hz). Furthermore, the full width at half maximum of the peaks in the power spectrum, $w\approx 0.5\ \mu$Hz,  corresponds to an e-folding lifetime  $1/(\pi w)$ of approximately 1 week or about 2--3 times the solar value \citep{Gizon03a}.

\begin{figure}[t]
\centering
\includegraphics[width=0.5\textwidth, trim=0.5cm 0 0.9cm 0, clip]{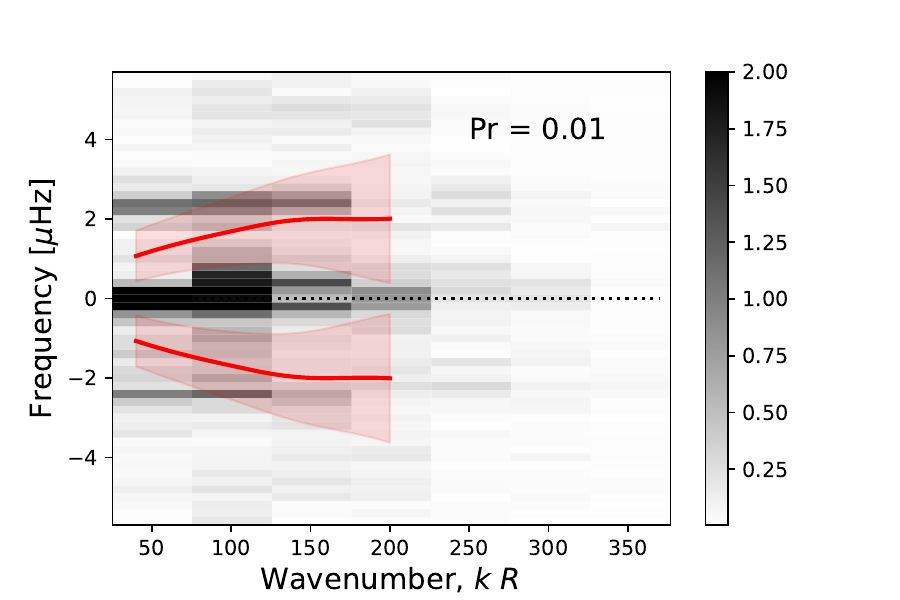} 
\caption{Two-dimensional power spectrum $P(k,\nu)$ of the horizontal velocity at height $z_0=0.7 H$, for the simulation with $\mathrm{Pr}=0.01$ (gray shades). To reduce random noise, the data were binned down to the spectral resolutions $\Delta k = 50/R$ and $\Delta \nu = 0.2\ \mu$Hz. The gray scale is saturated to highlight the peaks  near $\pm \nu_0$. For reference, the red curves correspond to the dispersion relation observed on the Sun \citep{Langfellner18a}, together with the full widths at half maximum of the peaks of power (red shades).
}
\label{fig3}
\end{figure}

The power spectrum versus wavenumber and frequency is shown as gray shades in Fig.~\ref{fig3} for the 
$\mathrm{Pr}=0.01$ simulation. The enhanced power near $k  \approx 50/R$ over a range of frequencies
is an indication of the typical convective cell size in the simulations, and is due to the fact that in two dimensions up- and downflow plumes may occasionally merge or split, but on average persist in their location and number. As reported on Fig.~\ref{fig3}, the oscillatory power seen in our simulations does not depend strongly on wave number for $kR<150$.
Overall, the comparison with the dispersion relation observed on the Sun at supergranular scales and larger \citep{Langfellner18a} is encouraging. 

To ensure that the phenomena described in this section are not dependent on choosing an intermediate 
value of the aspect ratio, we have repeated the calculation of the third simulation summarized in Table~1 
with its parameters $\mathrm{Ra}=8 \times 10^6$ and $\mathrm{Pr} = 0.025$ for an aspect ratio of $\Gamma=16$. 
This required doubling the number of grid points  along the horizontal direction to 6144. 
The simulation was  conducted over 10\% of the thermal diffusion time scale (1/3 of the extent of 
the run for $\Gamma=8$):  the time until a stable swaying oscillation sets in was found to almost double. 
Amplitudes and coupling appear less pronounced, but are still clearly detectable and appear in the correct
frequency range.

\section{Discussion}

Hard turbulence appears to be a necessary component to produce low-frequency oscillations in our setup (aspect ratio $L/H=8$). 
The low-viscosity regime ($\mathrm{Pr} \lesssim 0.01$) turns out to be essential to obtain a coupling of the dominant convective 
structures in time and space, via synchronized back-and-forth swaying oscillations. 
Such oscillations are unlikely to appear 
in traditional simulations of solar and stellar convection, typically carried out for Prandtl numbers of order unity,
as it is very difficult to achieve such low $\mathrm{Pr}$ in large-eddy simulations with realistic physics 
due to the unavoidable grid diffusivities \citep{Strugarek2016}.

As mentioned in the
introduction the phenomenon of flow reversals has been observed for setups resembling laboratory
experiments on convection (see also \citealt{Ahlers09}). 
Our model system has no solid vertical boundaries forcing the horizontal flow up- and 
downwards near the walls and we found no large scale circulation, as is also confirmed by the time 
averages of the horizontal flow being two orders of magnitudes smaller than horizontal root mean
square velocities and by the decaying of the former with time. 
However, the conceptual gap between 2D Boussinesq simulations and the solar case is not to be underestimated.
Whether swaying oscillations are preserved in three dimensions is an open question. One of the differences between turbulent flows
in 2D and 3D is the lack of vortex stretching for 2D flows (see Chap.~2.7.1 and 8.1 in \citealt{Lesieur08b}).
The most evident feature following from this restriction is that footpoints of up- and downflows at the bottom 
and at the top of the domain can merge or split in 2D. But they cannot move around each other and in this 
sense disappear which they can do in 3D. The role of additional constraining processes such as shear 
on the motions of up- and downflows has yet to be investigated. It has been suggested that scales larger than supergranulation are suppressed by rotation, which 
leads to a pronounced peak in power spectra of horizontal velocity \citep{Featherstone16} and constrains the 
inverse cascade of kinetic energy and thermal variance which limits the sizes of structures appearing 
in the flow \citep{Vieweg22}. 
The strong stratification of the solar convective zone breaks the symmetry 
present in RBC (see \citealt{Stein12} and references therein). 
One  consequence of stratification is the difference between the power spectra of horizontal and vertical velocities \citep{Lord2014}; as expected this is absent in our simulations.

In future work, additional physical ingredients may be studied in two dimensions, 
such as vertical gradients in density (stratification) and in velocity (due to shear).
The latter will break the east-west symmetry and cause 
the pattern to propagate in a preferred direction (on the Sun, the pattern is observed to propagate in the 
prograde direction). To avoid some of the limitations of the present study requires 3D flow simulations.
These should be investigated in later, follow-up work. Unfortunately, 3D DNS simulations are very computer 
intensive \citep{Pandey2022}, as we require hundreds of convective turnovers to identify a phenomenon such 
as the swaying oscillations reported in this paper.

Helioseismology offers a way forward to determine observationally whether the evolution of the velocity field in the Sun resembles the picture of the swaying oscillations  sketched in our 2D simulations. For this purpose, maps of  horizontal flows at different times and depths across the supergranulation layer can in principle be obtained \citep{Gizon2010} and compared with the model\footnote{See movie for $u_x$ at the following DOI: \href{https://doi.org/10.17617/3.BHN9A4}{10.17617/3.BHN9A4}.}.  
The evolution of the observed intensity contrast \citep{Langfellner16} may provide an additional physical  constraint on  models.

Finally, we wish to highlight that the low-Pr-number simulations presented here provide insights of broad relevance. They reveal that interesting and unexpected regimes 
of Rayleigh-B\'enard convection remain to be explored, regimes that are currently inaccessible to laboratory experiments.

\begin{acknowledgements}
L.G. proposed the project. F.K. designed the model system. F.K., D.F. and F.Z. performed research. F.K., D.F., D.K. and L.G. analyzed data. F.K., D.F. and L.G. wrote the article.
We thank Ambrish Pandey for useful discussions. The Austrian Science Fund (FWF) provided support under projects P~29172-N (F.K., D.F. and D.K.), 
P~33140-N (F.K. and D.F.), and P~35485-N (F.K. and D.F.). F.K.\ and D.F.\ are thankful for the hospitality of the Wolfgang Pauli Institute, Vienna, and acknowledge
the support of the Faculty of Mathematics at the University of Vienna by providing them with a Senior Research Fellow status.
L.G. acknowledges support from European Research Council (ERC) Synergy Grant WHOLESUN \#810218. 
F.Z. acknowledges support by German Aerospace Center (DLR) grants  50WK2270A (DAIMLER), 50WM2163 (AID), and 50WM2354 (AID2).
Computing time was provided by the North-German Supercomputing Alliance (HLRN). Local computing resources were funded by the 
Land of Niedersachsen and the German Aerospace Center (German Data Center for SDO) through grants to L.G. The Vienna Scientific Cluster (VSC)
has been used for development work preparing the numerical simulations presented here. D.K. was a member of the International Max Planck Research School 
for Solar System Science at the University of G\"ottingen.
\end{acknowledgements}

\bibliographystyle{aa}
\bibliography{references} 

\end{document}